\documentclass[%
 amsmath,amssymb,
 aps,
 twocolumn,
prb,
citeautoscript
]{revtex4-2}

\setcitestyle{super}

\usepackage[caption=false]{subfig}

\usepackage{graphicx}
\usepackage{dcolumn}
\usepackage{bm}
\usepackage{longtable}
\usepackage{float}
\usepackage{hyperref}
\usepackage[version=3]{mhchem}
\usepackage{color}

\newcolumntype{C}[1]{>{\centering\arraybackslash}m{#1}}

\begin{document}

\preprint{APS/123-QED}

\title{Symmetry breaking slows convergence of the ADAPT Variational Quantum Eigensolver}

\author{Luke W. Bertels}
\author{Harper R. Grimsley}
\affiliation{Department of Chemistry, Virginia Tech, Blacksburg, VA 24061, USA}
\author{Sophia E. Economou}
\author{Edwin Barnes}
\affiliation{Department of Physics, Virginia Tech, Blacksburg, VA 24061, USA}
\author{Nicholas J. Mayhall}
\email{nmayhall@vt.edu}
\affiliation{Department of Chemistry, Virginia Tech, Blacksburg, VA 24061, USA}

\date{\today}

\begin{abstract}
Because quantum simulation of molecular systems is expected to provide the strongest advantage over classical computing methods for systems exhibiting strong electron correlation, it is critical that the performance of VQEs be assessed for strongly correlated systems. 
For classical simulation, strong correlation often results in symmetry-breaking of the Hartree-Fock reference, leading to 
L\"owdin's well-known ``symmetry dilemma'' whereby accuracy in the energy can be increased by breaking spin or spatial symmetries. 
Here, we explore the impact of symmetry breaking on the performance of ADAPT-VQE using 
two strongly correlated systems: (i) the ``fermionized" anisotropic Heisenberg model, where the anisotropy parameter controls the correlation in the system, and (ii) symmetrically-stretched linear \ce{H4}, where correlation increases with increasing \ce{H}-\ce{H} separation. 
In both of these cases, increasing the level of correlation of the system leads to spontaneous symmetry breaking (parity and $\hat{S}^{2}$, respectively) of the mean-field solutions.
We analyze the role that symmetry breaking in the reference states and orbital mappings of the fermionic Hamiltonians have on the compactness and performance of ADAPT-VQE. 
We observe that improving the energy of the reference states by breaking symmetry has a deleterious effect on ADAPT-VQE by increasing the length of the ansatz necessary for energy convergence and exacerbating the problem of ``gradient troughs".  
\end{abstract}

\maketitle


\section{\label{sec:intro}Introduction}
The simulation of ground electronic states of molecular Hamiltonians is a fundamental goal of theoretical chemistry. 
Several classes of methods exist within the realm of classical computing for treating these systems, including density functional theory (DFT)\cite{hohenberg1964inhomogeneous,kohn1965self}, Hartree-Fock (HF) theory\cite{fock1930hf,dirac1930note}, M{\o}ller-Plesset (MP) perturbation theory\cite{moller1934note}, coupled cluster (CC) theory\cite{coester1960short,cizek1966correlation}, configuration interaction (CI), and many others. 
While DFT and HF are by nature approximate, MP2, CC, and CI are systematically improvable.
For systems with weakly correlated electrons, inclusion of only single and double excitations in MP or CC theory is sufficient for accurate results, while for systems with strongly correlated electrons, low excitation rank methods fail to capture the physics required to provide accurate energies, although increasing the excitation rank leads to a rapid increase in the computational resources required, with exact treatment [full CI (FCI)] requiring combinatorial scaling with system size to cover the Hilbert space of the system. 

Simulation of chemical systems on quantum computers, however, offers an attractive alternative to classical simulations, as the quantum mechanical structure is efficiently captured in the quantum nature of the device\cite{feynman1982simulating}, i.e., the combinatorial growth of the Hilbert space with system size is absorbed by the quantum processor\cite{aspuru2005simulated}. 
The promise of quantum computation for chemistry is currently limited by the small numbers of qubits ($10^{1}-10^{2}$) in existing quantum devices and the quality of those qubits (limited coherence times). 
In this noisy intermediate scale quantum (NISQ) era, the success of quantum simulation depends on both the quality of qubits and the ability of quantum algorithms to cope with these limitations\cite{preskill2018quantum}.    


The Variational Quantum Eigensolver (VQE), originally proposed by Peruzzo et. al.\cite{peruzzo2014variational}, offers an approach for quantum simulation of chemical Hamiltonians. 
VQE is a hybrid quantum-classical algorithm in which the computational work is divided between a quantum processor and a classical co-processor\cite{mcclean2016theory}. 
In this scheme, one prepares parameterized trial states $|\psi(\vec{\theta})\rangle$ on the quantum processor and minimizes the expectation value of the Hamiltonian with respect to the ansatz parameters:
\begin{eqnarray}
    E &=& \min_{\vec{\theta}}\langle \psi(\vec{\theta})|\hat{H}|\psi(\vec{\theta})\rangle \label{eqn:e_vqe1} \\
    &=& \min_{\vec{\theta}}\sum_{i} g_{i}\langle\psi(\vec{\theta})|\hat{o}_{i}|\psi(\vec{\theta})\rangle,\label{eqn:e_vqe2}
\end{eqnarray}
where $g_{i}$ are the classically precomputed one- and two-electron integrals, and $\hat{o}_{i}$ are the corresponding one- and two-electron operators. 
As the different $\hat{o}_{i}$ terms do not generally commute, state preparation and measurement of the terms in Eq.~\ref{eqn:e_vqe2} must be performed multiple times to statistically converge the expectation values.  

Here, the quantum computer is used to prepare trial states and measure the molecular Hamiltonian, while the classical computer is used to determine ansatz parameter updates. 
Trial states are prepared by applying a parameterized unitary operator, $\hat{U}(\vec{\theta})$, to a reference state $|\psi^{(0)}\rangle$: 
\begin{equation}\label{eqn:vqe_wfn}
    |\psi(\vec{\theta})\rangle = \hat{U}(\vec{\theta})|\psi^{(0)}\rangle.
\end{equation}
Several VQE ans\"atze have been explored for theoretical studies\cite{peruzzo2014variational,omalley2016scalable,mcclean2016theory,mcclean2017hybrid,barkoutsos2018quantum,romero2018strategies,colless2018computation,lee2018generalized,dallaire2019low,ryabinkin2018qubit} and on quantum hardware\cite{omalley2016scalable,colless2018computation, kandala2017hardware,shen2017quantum,hempel2018quantum}, many of which are modifications of the unitary coupled cluster (UCC) ansatz\cite{bartlett1989alternative, kutzelnigg1991error,taube2006new,harsha2018difference} from classical electronic structure theory. 
While in principle a circuit implementation of an arbitrarily expressive ansatz can map the reference to any state in the Hilbert space (including the exact FCI state), practical limits on the depth of circuits that may be implemented within the coherence times of NISQ devices imposes limits on the structure of $U(\vec{\theta})$. 
Therefore, while the reduced circuit depth of VQEs versus PEA (at the cost of many more measurements) makes VQEs attractive for NISQ devices, the limits imposed on $\hat{U}(\vec{\theta})$ mean that VQEs typically produce approximate solutions. 
The accuracy of VQEs is, therefore, ultimately limited by the variational flexibility of the predefined ansatz. 

Unlike VQEs with statically defined ans\"atze, the adaptive problem-tailored VQE (ADAPT-VQE) method, developed by Grimsley et al.\cite{grimsley2019adaptive}, avoids a predefined unitary ansatz by constructing an arbitrarily accurate quasi-optimal ansatz on the fly. 
This is achieved by iteratively growing the ansatz by adding operators from a pool one-at-a-time as informed by the Hamiltonian. 
ADAPT-VQE has been shown to simultaneously provide smaller gate counts and errors than traditional VQE methods\cite{grimsley2019adaptive,claudino2020benchmarking}. Further improvements on the ADAPT-VQE framework have come from the introduction of qubit-based operator pools (qubit-ADAPT-VQE)\cite{tang2021qubit} and minimally complete pools to reduce measurement overhead\cite{shkolnikov2021avoiding}. 
The success of ADAPT-VQE has inspired the development of other adaptive VQEs as well, including iterative qubit excitation based VQE (QEB-ADAPT-VQE)\cite{yordanov2021qubit}, mutual information-assisted adaptive VQE\cite{zhang2021mutual}, and the adaptive variational quantum imaginary time evolution (AVQITE) method\cite{gomes2021adaptive}. 

An additional motivation for the adaptive construction of ans{\"a}tze is the ability to adapt to systems that are strongly correlated, where the performance of classical methods and even traditional VQEs is expected to suffer. 
In this work, we investigate the performance of ADAPT-VQE on two distinct systems that display variable amounts of strong correlation: 
(i) the fermionized anisotropic Heisenberg model, where the anisotropy parameter allows for control over the level of correlation in the system, 
and (ii) the symmetric dissociation of linear \ce{H4}.
In both of these cases, increasing the level of correlation of the system leads to spontaneous symmetry breaking (parity and $\hat{S}^{2}$, respectively) of the mean-field solutions.
We explore the roles played by these symmetries, both in the reference state and the operator pool, for ADAPT-VQE, highlighting their importance in generating compact ans\"{a}tze and preventing premature convergence of the algorithm. 
Our results bolster the findings of Barron et al.\cite{barron2021preserving} and Shkolnikov et al.\cite{shkolnikov2021avoiding} on the importance of building the symmetries of the Hamiltonian into the operator pools.

\section{Background}
\subsection{\label{sec:adapt-vqe}ADAPT-VQE Algorithm}
Unlike traditional VQE, which begins with a predetermined form of the unitary $U(\vec{\theta})$, ADAPT-VQE iteratively grows a problem-tailored unitary ansatz by adding operators one-at-a-time from a predetermined operator pool. 
Before the algorithm begins, the Hamiltonian coefficients are computed and mapped to a qubit representation, as in traditional VQE.
The operators, $\hat{A}_k$, in the pools used in this work take the form of anti-Hermitian sums of generalized excitation and de-excitation operators, e.g.,
\begin{eqnarray}
    \hat{A}^{p}_{q} &=& \hat{t}^{p}_{q} -  \hat{t}^{q}_{p}, \nonumber \\
    &=& t^{p}_{q}\left(\hat{a}^{\dagger}_{p}\hat{a}_{q} - \hat{a}^{\dagger}_{q}\hat{a}_{p}\right), \\
    \hat{A}^{pq}_{rs} &=& \hat{t}^{pq}_{rs} - \hat{t}^{rs}_{pq}, \nonumber \\
    &=& t^{pq}_{rs}\left(\hat{a}^{\dagger}_{p}\hat{a}^{\dagger}_{q}\hat{a}_{s}\hat{a}_{r} - \hat{a}^{\dagger}_{r}\hat{a}^{\dagger}_{s}\hat{a}_{q}\hat{a}_{p}\right),
\end{eqnarray}
where $p$, $q$, $r$, and $s$ are arbitrary spin-orbital indices. 
Exponentiation of these anti-Hermitian operators yields unitary operators. 
While other operator pools have been explored,\cite{yordanov2020efficient,tang2021qubit,yordanov2021qubit,shkolnikov2021avoiding} we focus here on the fermionic operator pool.
The ADAPT-VQE trial state is then initialized with a reference state that is easily prepared on the device, typically a product state corresponding to the HF determinant. 
To grow the ansatz, the current trial state, $|\psi^{(n)}\rangle$, is prepared on the device and the gradient of the energy with respect to the operator parameters $\theta_{k}$ for each operator $\hat{A}_{k}$ in the pool is measured. 
This is done by measuring the expectation value of commutator of the Hamiltonian and the operators for the current state:
\begin{equation}\label{eqn:adapt_grad}
    \frac{\partial E^{(n)}}{\partial \theta_{k}} = \left\langle\psi^{(n)}\left|\left[\hat{H},\hat{A}_{k}\right]\right|\psi^{(n)}\right\rangle.
\end{equation}
This gradient measurement step of ADAPT-VQE is highly parallelizable over multiple uncoupled devices. 
The operator corresponding to the largest gradient magnitude is then used to form the new trial state ansatz:
\begin{eqnarray}
    |\psi^{(n+1)}(\vec{\theta}^{(n+1)})\rangle &=& e^{\theta_{n+1}\hat{A}_{n+1}}|\psi^{(n)}\rangle \nonumber\\
    &=& e^{\theta_{n+1}\hat{A}_{n+1}}e^{\theta_{n}\hat{A}_{n}}\cdots e^{\theta_{1}\hat{A}_{1}}|\psi^{(0)}\rangle.
\end{eqnarray}
The new parameter $\theta_{n+1}$ is initialized to 0, while the initial values for the other parameters are taken to be the optimized values from the previous iteration. 
The new ansatz is then optimized over all $\vec{\theta}^{(n+1)}$ via a VQE subroutine to yield $|\psi^{(n+1)}\rangle$. 
From here the algorithm repeats by returning to the operator gradient measurement step. 
Convergence of the ADAPT-VQE algorithm may be determined in a number of ways, including the norm ($l^2$ or $l^\infty$) of the operator gradient $\left|\frac{\partial E^{(n)}}{\partial \theta_{k}} \right| < \epsilon$, the variance of the ADAPT-VQE state $\langle\psi^{(n)}|\hat{H}^2|\psi^{(n)}\rangle - (E^{(n)})^{2} < \epsilon$, and the energy change between iterations $|E^{(n)} - E^{(n-1)}| < \epsilon$. 
The operator pools are not ``drained" by the addition of an operator to the ansatz; a given operator may be added to the ansatz more than once, with different parameters for each occurrence. 
Because of this, ADAPT-VQE can be viewed as an algorithm that approximates the exact (FCI) ground state to arbitrary accuracy by appending multiple instances of the operators: 
\begin{equation}
    |\psi^{FCI}\rangle = \prod_{l}^{\infty}\prod_{k}e^{\theta^{(l)}_{k}\hat{A}_{k}}|\psi^{(0)}\rangle,
\end{equation}
where the parameters $\theta^{(l)}_{k}$ are allowed to vary independently for different $l$. 
Ref.~\citenum{grimsley2019adaptive} presents a more detailed explanation and demonstration of ADAPT-VQE and this connection to FCI. Similar work on the exactness of general trotterized UCC variants has been explored by Evangelista et al.\cite{evangelista2019exact}  

\subsection{\label{sec:Heisenberg}The Heisenberg Model}
While spin Hamiltonians are most often associated with condensed matter physics, these Hamiltonians are also often used in the context of chemistry as model systems to develop a coarse-grained understanding of certain molecular interactions.\cite{calzado2002analysis, maurice2009universal, monari2010determination,malrieu2014magnetic, mayhallComputationalQuantumChemistry2014, mayhallComputationalQuantumChemistry2015a, mayhallModelHamiltoniansInitio2016a, coen2016magnetic,abrahamSimpleRulePredict2017a, pokhilko2020effective, kotaru2022magnetic}
Such models are useful for describing the interactions between open-shell fragments, such as metal atoms in multi-metal organometallic complexes. 
These interactions are broadly classified as ferromagnetic coupling or antiferromagnetic coupling based on whether or not the ground state has a spin magnetic moment. 

In this picture, the unpaired electrons on a given metal atom are aligned parallel to each other while the unpaired electrons on different metal atoms align either parallel (ferromagnetic coupling) or antiparallel (antiferomagnetic coupling) to each other. 
While purely \textit{ab initio} approaches to describe these systems must contend with strongly interacting electrons within nearly degenerate orbitals, the effective-Hamiltonian approaches reduce these to the interactions between the net spins on different fragments. 
The exchange interaction, a consequence of Fermi statistics, 
provides the energetic driving force behind this coupling. 
For fixed oxidation states, the Heisenberg-Dirac-van Vleck Hamiltonian (HDvV)\cite{heisenberg1928theorie,dirac1926theory,vanvleck1932theory} provides a simple model that depends on the net spin of the different metal centers:
\begin{eqnarray}
    \hat{H}^{\text{HDvV}} &=& -2 \sum_{ij}J_{ij} \hat{\vec{S}}_{i} \cdot \hat{\vec{S}}_{j} \label{eqn:hdvv1}\\
    &=& -2 \sum_{ij}J_{ij} \left(\hat{S}^{x}_{i}\hat{S}^{x}_{j} + \hat{S}^{y}_{i}\hat{S}^{y}_{j} + \hat{S}^{z}_{i}\hat{S}^{z}_{j}\right). \label{eqn:hdvv2}
\end{eqnarray}
$J_{ij} > 0$ couples sites $i$ and $j$ ferromagnetically while $J_{ij} < 0$ couples them antiferromagnetically. 
The problem then shifts (slightly) from describing the many interactions between the electrons to describing the interactions between the spins, namely obtaining values for $J_{ij}$. 

The Heisenberg spin Hamiltonian can also be considered as a model for strong fermionic correlation as well, when viewed in the strong-correlation limit of the fermionic Hubbard model. 
The fermionic Hubbard Hamiltonian is given as 
\begin{equation}
    \hat{H}^{\text{Hubbard}} = t\sum_{ij}\sum_{\sigma} \hat{a}^{\dagger}_{i,\sigma}\hat{a}_{j,\sigma} + \frac{1}{2}U\sum_{i}\hat{n}_{i,\sigma}\hat{n}_{i,\bar{\sigma}},
\end{equation}
where $t$ is the single-electron hopping, and $U$ is the two-electron repulsion. 
When $\frac{t}{U} \gg 1$, the system is dominated by hopping (kinetic energy-like), and in the limit $U\rightarrow0$ it becomes a free-electron system where the electrons delocalize over the entire lattice. 
In this regime the delocalized state may be taken as the zeroth-order solution, with correlations between electrons handled by perturbation theory. 
In the opposite limit, where $\frac{t}{U} \ll 1$, the system becomes localized. In this regime, degenerate perturbation theory may be used to treat delocalization as a perturbation to the $\binom{N}{k}$-fold degenerate localized ground states, where $N$ is the number of sites and $k$ the number of electrons.  
This approach yields the Heisenberg Hamiltonian and at second order $J^{(2)}_{ij} = -\frac{t^{2}}{U}$ (see derivation in Ref. \citenum{cleveland1976obtaining}). 
This connection between the Hubbard model in the limit of large electron-electron repulsion and the Heisenberg spin Hamiltonian suggests the latter as a model for studying strong correlation. 
While our use of the Heisenberg model is as a proxy for chemical systems, the use of quantum computers to simulate model Hamiltonians is an important field in its own right, with much recent work in this context\cite{dallaire2016method,cai2020resource,cade2020strategies,bilkis2021semi,vandyke2021preparing,vandyke2022preparing,selvarajan2021variational,gyawali2022adaptive,jattana2022assessment,barron2021preserving}.

\subsubsection{Anisotropic Heisenberg Hamiltonian}
The HDvV Hamiltonian given in Eqs.~\eqref{eqn:hdvv1} and~\eqref{eqn:hdvv2} is referred to as being \emph{isotropic}, meaning that the $x$, $y$, and $z$ components of the total spin are treated equivalently. 
If interactions are present that break this equivalence (e.g. dipolar-like couplings), the resulting effective spin Hamiltonian becomes \emph{anisotropic}. 
The anisotropic Heisenberg model, also known as the XXZ model, has a Hamiltonian given by 
\begin{equation}\label{eqn:xxz_ham}
 \hat{H}^{\text{aniso}} = -2J\sum_{\langle ij\rangle} \left(\hat{S}^{x}_{i}\hat{S}^{x}_{j} + \hat{S}^{y}_{i}\hat{S}^{y}_{j}\right) - 2K \sum_{\langle ij\rangle}\hat{S}^{z}_{i}\hat{S}^{z}_{j},
\end{equation}
where $\langle ij\rangle$ restricts the sum over nearest-neighbor sites. 

\subsubsection{Fermionization of 1D spin Hamiltonians}
While application of degenerate perturbation theory to the Hubbard model in the $\frac{t}{U} \ll 1$ limit to yield the Heisenberg Hamiltonian provides a connection between fermions and spins, this connection is made more general by the Jordan-Wigner (JW) transform\cite{wigner1928paulische}. 
The JW transform has become ubiquitous in quantum simulation of chemistry Hamiltonians as a means to map fermionic Hamiltonians onto qubit Hamiltonians; however, the transformation was originally proposed as a means to map spins onto fermions. 
For a one-dimensional (1D) spin-1/2 lattice, the anisotropic Heisenberg Hamiltonian is ``fermionized" by first writing Eq.~\ref{eqn:xxz_ham} in terms of Pauli ladder operators and then substituting them with fermionic creation/annihilation operators:
\begin{align}
    \hat{H}^{\text{aniso}} = -\frac{J}{2} \sum_{i}& \left(\hat{\sigma}^{x}_{i}\hat{\sigma}^{x}_{i+1} + \hat{\sigma}^{y}_{i}\hat{\sigma}^{y}_{i+1}\right) \nonumber \\
    -\frac{K}{2}\sum_{i}&\hat{\sigma}^{z}_{i}\hat{\sigma}^{z}_{i+1} \\
    =-J \sum_{i}& \left(\hat{\sigma}^{+}_{i}\hat{\sigma}^{-}_{i+1} + \hat{\sigma}^{-}_{i}\hat{\sigma}^{+}_{i+1}\right) \nonumber \\
    -\frac{K}{2}\sum_{i}&\hat{\sigma}^{z}_{i}\hat{\sigma}^{z}_{i+1} \\
    = -J \sum_{i}& \left(\hat{a}^{\dagger}_{i}\hat{a}_{i+1} + \hat{a}_{i}\hat{a}^{\dagger}_{i+1}\right) \nonumber \\
    -K \sum_{i}& \left(2\hat{a}^{\dagger}_{i}\hat{a}_{i}\hat{a}^{\dagger}_{i+1}\hat{a}_{i+1} - \hat{a}^{\dagger}_{i}\hat{a}_{i} \right.\nonumber \\ &\left.-\hat{a}^{\dagger}_{i+1}\hat{a}_{i+1} + \frac{1}{2} \right) .
\end{align}
For $K=0$, this fermionized Hamiltonian reduces to a one-electron Hamiltonian that is easily diagonalized to yield non-interacting fermions. 
This model is known as the XY model, and despite acting on the full Hilbert space in the spin representation, it has a trivially simple solution in the fermionic representation. 
Table ~\ref{tab:xxz_lims} summarizes the different limits of the anisotropic Heisenberg Hamiltonian. The ratio of $K/J$ is seen as a correlating parameter, as below the isotropic point ($K/J = 1$) increasing $K/J$ increases the correlation in the system by encouraging localization. 
Above the isotropic point this correlation decreases with increasing $K/J$. Beginning at the isotropic point and for all larger $K/J$, the mean-field solution is seen to break spatial (parity) symmetry.  
In the first set of results, we apply ADAPT-VQE to the fermionized anisotropic Heisenberg model to investigate the role of parity symmetry in the performance of ADAPT-VQE for both local and non-local representations.
\begin{table*}
\caption{\label{tab:xxz_lims}Different cases of the Hamiltonian in Eq.~\eqref{eqn:xxz_ham} and comparisons to fermionic models when fermionized.}
\begin{ruledtabular}
\begin{tabular}{p{0.10\linewidth}p{0.40\linewidth}p{0.45\linewidth}}
Case & Hamiltonian & Features \\
\hline
$K=0$ & $\hat{H}^{\text{XY}} = -2J \sum\limits_{\langle ij\rangle}\left(\hat{S}^{x}_{i}\hat{S}^{x}_{j} + \hat{S}^{y}_{i}\hat{S}^{y}_{j}\right)$ & Free fermion model, completely delocalized, no entanglement in the ground state \\
$K=J$ & $\hat{H}^{\text{HDvV}} = -2J \sum\limits_{\langle ij\rangle}\left(\hat{S}^{x}_{i}\hat{S}^{x}_{j} + \hat{S}^{y}_{i}\hat{S}^{y}_{j} + \hat{S}^{z}_{i}\hat{S}^{z}_{j}\right)$ & Isotropic, competition between localization and delocalization, entangled ground state \\
$J=0$ & $\hat{H}^{\text{Ising}} = -2K \sum\limits_{\langle ij\rangle}\hat{S}^{z}_{i}\hat{S}^{z}_{j}$ & Degenerate N\'{e}el ground states, completely localized, no entanglement in ground state  
\end{tabular}
\end{ruledtabular}
\end{table*}

\section{Computational Details}
We employ an antiferromagnetically coupled ($J<0$; $K<0$) anisotropic Heisenberg Hamiltonian on a 1D, eight-site lattice as a model system for our calculations. 
Calculations are performed within the $M_{s}=0$ space, which after JW transformation yields a four-fermion system (half-filling). 
We survey correlation parameter values $K/J$ ranging from 0.001 to 100. The OpenFermion\cite{mcclean2020openfermion} electronic structure theory package is used to build the nearest-neighbor (local) spin Hamiltonians and transform them into local fermionic Hamiltonians.\footnote{The locality of the spin Hamiltonian in the fermionic representation arises only from the fact that the spin Hamiltonian is local and 1D, which cancels all the $Z$ strings that enforce antisymmetry.} 
These local fermionic Hamiltonians are diagonalized to yield full configuration interaction (FCI) energies and wavefunctions. 
For each surveyed value of $K/J$, the ground state wavefunction has even symmetry (\textit{gerade}, $g$) with respect to the lattice, and the first excited state has odd symmetry (\textit{ungerade}, $u$). 
These two states become degenerate in the $K/J\rightarrow\infty$ limit. 
The local Hamiltonians are then read into the PySCF\cite{sun2018pyscf,sun2020recent} electronic structure theory package to perform HF. 
For $K/J < 1$, a wavefunction stability analysis of the HF solutions confirms that the stable HF solution has an even symmetry with respect to the center of the lattice. 
We follow this solution for $K/J \geq 1$ to yield symmetry-preserving HF solutions; however, performing a stability analysis on these solutions yields a more stable broken-symmetry solution. 
The ``molecular" orbitals (MOs) from both the symmetry-preserving and symmetry-broken HF solutions are then used to transform the local Hamiltonians into non-local MO bases. 

\begin{figure}
\includegraphics[width=\linewidth]{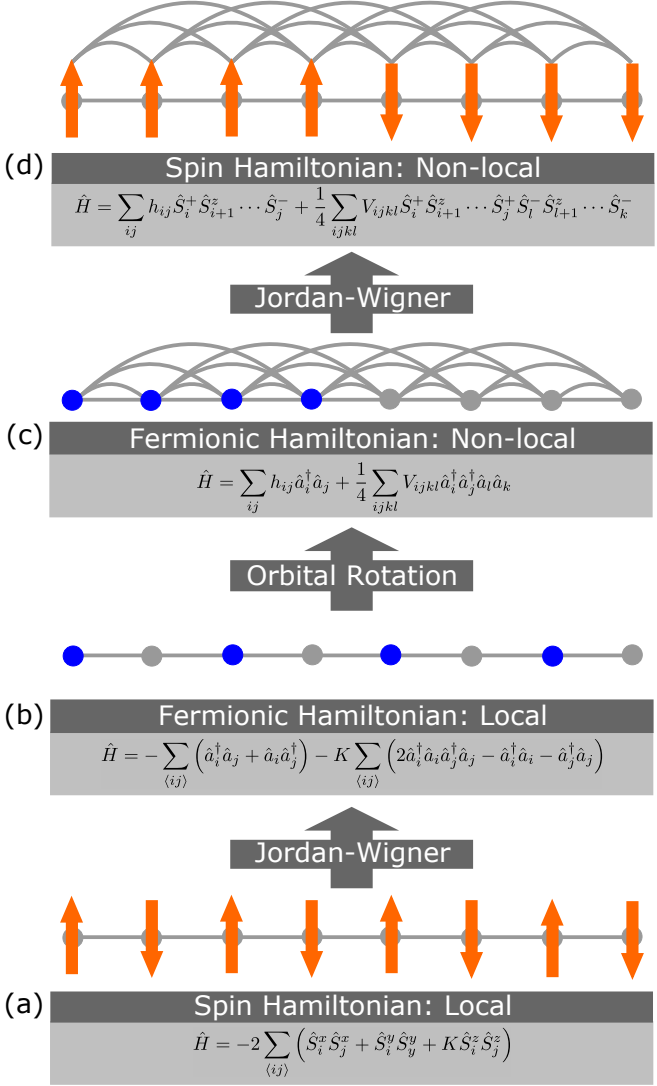} \\
\caption{\label{fig:jordanwigner} Depiction of the use of the Jordan-Wigner transformation and Hartree-Fock to transform the anisotropic Heisenberg Hamiltonian from a local representation to a non-local representation. Grey lines between points depict interactions (not all are shown, but meant to represent the locality/non-locality of the system). Constant terms are neglected and $J$ is set to 1.  }
\end{figure}

We perform second-order M{\o}ller-Plesset perturbation theory (MP2)\cite{moller1934note}, and traditional (non-unitary) coupled cluster theory with single and double excitations (CCSD)\cite{purvis1982full} on top of these HF solutions with PySCF. 
The non-local fermionic Hamiltonians are then translated into non-local qubit Hamiltonians within the ADAPT-VQE procedure. 
This process of transforming from local spin Hamiltonians to non-local spin Hamiltonians is illustrated in Fig.~\ref{fig:jordanwigner}. 

For linear \ce{H4}, we survey \ce{H}--\ce{H} separations from  0.50\,\AA{} to 3.00\,\AA. 
Classical electronic structure calculations are performed with PySCF, and the STO-3G minimal basis\cite{hehre1969self} is used for both the classical methods and ADAPT-VQE simulations.

The ADAPT-VQE calculations are simulated without noise on a development branch of our in-house code\cite{adaptcode}, 
which in turn uses OpenFermion\cite{mcclean2020openfermion} for the JW operator transformations and SciPy\cite{virtanen2020scipy} for BFGS optimization of the parameters in the VQE subroutine. 

For the fermionized, anisotropic Heisenberg Hamiltonians, we perform ADAPT-VQE simulations with five combinations of reference states and orbital bases (used to transform the Hamiltonians and define the fermionic operator pools) to investigate the roles that symmetry plays in ADAPT-VQE as correlation increases. 
\begin{enumerate}
\item \textbf{Symmetry-preserving HF orbitals and reference state}: The canonical HF orbitals are used to transform the Hamiltonians from the local site orbital basis to the nonlocal, symmetry-preserving MO bases. 
When the parity of the system is enforced at the HF level, these canonical orbitals are of either $g$ or $u$ character, and as such the determinants are eigenstates of the parity operator with parities determined by the occupied orbitals. 
Similarly, the parities of the pool operators are determined by the orbitals used to define the excitations/de-excitations.
The reference state is the JW transformed HF state,
\begin{equation}\label{eqn:hf_ref}
    |\psi^{(0)}_{\text{HF}}\rangle = |11110000\rangle,
\end{equation}
which has $g$ symmetry,
and where orbitals are ordered in increasing energy from left to right. 
This reference state is the exact ground state in the free-fermion limit ($K\rightarrow0$).  
\item \textbf{Symmetry-breaking HF orbitals and reference state:} The canonical HF orbitals are used to transform the Hamiltonian from the local site basis to the nonlocal, broken-symmetry HF orbitals. 
With the onset of symmetry-breaking, these canonical orbitals are of neither $g$ nor $u$ character, and therefore neither the determinants nor operators have parity symmetry. 
The reference state is the JW transformed HF state (Eq.~\ref{eqn:hf_ref})
which also breaks parity symmetry due to the symmetry-breaking of the underlying orbitals. 
\item \textbf{Local orbital basis, N\'{e}el reference state}: The Hamiltonian is expressed in the local site basis. 
As the site orbital basis has no parity symmetry, the determinant basis and operators lack parity symmetry. 
The reference state is the N\'{e}el state,
\begin{equation}\label{eqn:neel_ref}
    |\psi^{(0)}_{\text{N\'{e}el}}\rangle = |10101010\rangle,
\end{equation}
which does not have parity symmetry. 
This reference state is energetically exact, though symmetry-broken, in the Ising limit ($J\rightarrow0$), analogous to how unrestricted HF becomes exact for separated hydrogen atoms.
\item \textbf{Local orbital basis, cat$_{+}$ reference state}: The Hamiltonian is expressed in the local site orbital basis. 
While the basis and operators have no parity symmetry due to the asymmetry of the site orbital basis, the cat$_{+}$ state, given as the plus superposition of the two complementary N\'{e}el states,
\begin{equation}\label{eqn:sc+_ref}
    |\psi^{(0)}_{\text{cat}_{+}}\rangle = \frac{1}{\sqrt{2}}\left(|10101010\rangle + |01010101\rangle\right),
\end{equation}
is used as the reference state and has $g$ symmetry.
This reference state lies in the two-fold degenerate subspace of the exact ground state in the Ising limit ($J\rightarrow0$).
\item \textbf{SALC orbital basis, cat$_{+}$ reference state}: A symmetry-adapted basis formed by taking the plus and minus linear combinations of complementary site orbitals is used to transform the Hamiltonians from the site orbital basis to the SALC orbital basis. 
By construction, these SALC orbitals are of either $g$ or $u$ character, and as such the determinant basis has parity symmetry. 
Similarly, the parities of the operators are determined by the orbitals used to define the excitations/de-excitations. 
The reference state is the cat$_{+}$ state in the SALC orbital basis,
\begin{align}\label{eqn:salc_sc+_ref}
    |\psi^{(0)}_{\text{SALC,cat}_{+}}\rangle = \frac{1}{\sqrt{2}}&\left(|10101010\rangle + |01100110\rangle \nonumber \right. \\
    &+ |10011001\rangle + |01010101\rangle \nonumber \\
    &-|01101001\rangle - |10100101\rangle \nonumber \\
    &\left.- |01011010\rangle - |10010110\rangle\right),
\end{align}
which has $g$ symmetry. 
\end{enumerate}
For all five combinations of orbital bases and reference states, we use a fermionic generalized singles and doubles (GSD) operator pool without symmetry adaptation of the fermionic operators. 

For the symmetric dissociation of \ce{H4}, we compare the performance of ADAPT-VQE
when performed with spin-restricted HF (rHF)
and spin-unrestricted HF (uHF) orbitals. 
This allows us to determine the impact of symmetry breaking in the representation. 
For rHF we further explore the impact of spin-adapting the operator pool, 
by using both the singlet-GSD (sGSD) pool, where the pool operators are symmetry-adapted linear combinations of excitation/de-excitation operators, and the unrestricted-GSD (uGSD) pool, where the excitation/de-excitation operators in the operator pool are not symmetry-adapted.

\section{Results}
\subsection{Anisotropic Heisenberg Model}

\begin{figure*}
     \centering
     \includegraphics[width=.45\linewidth]{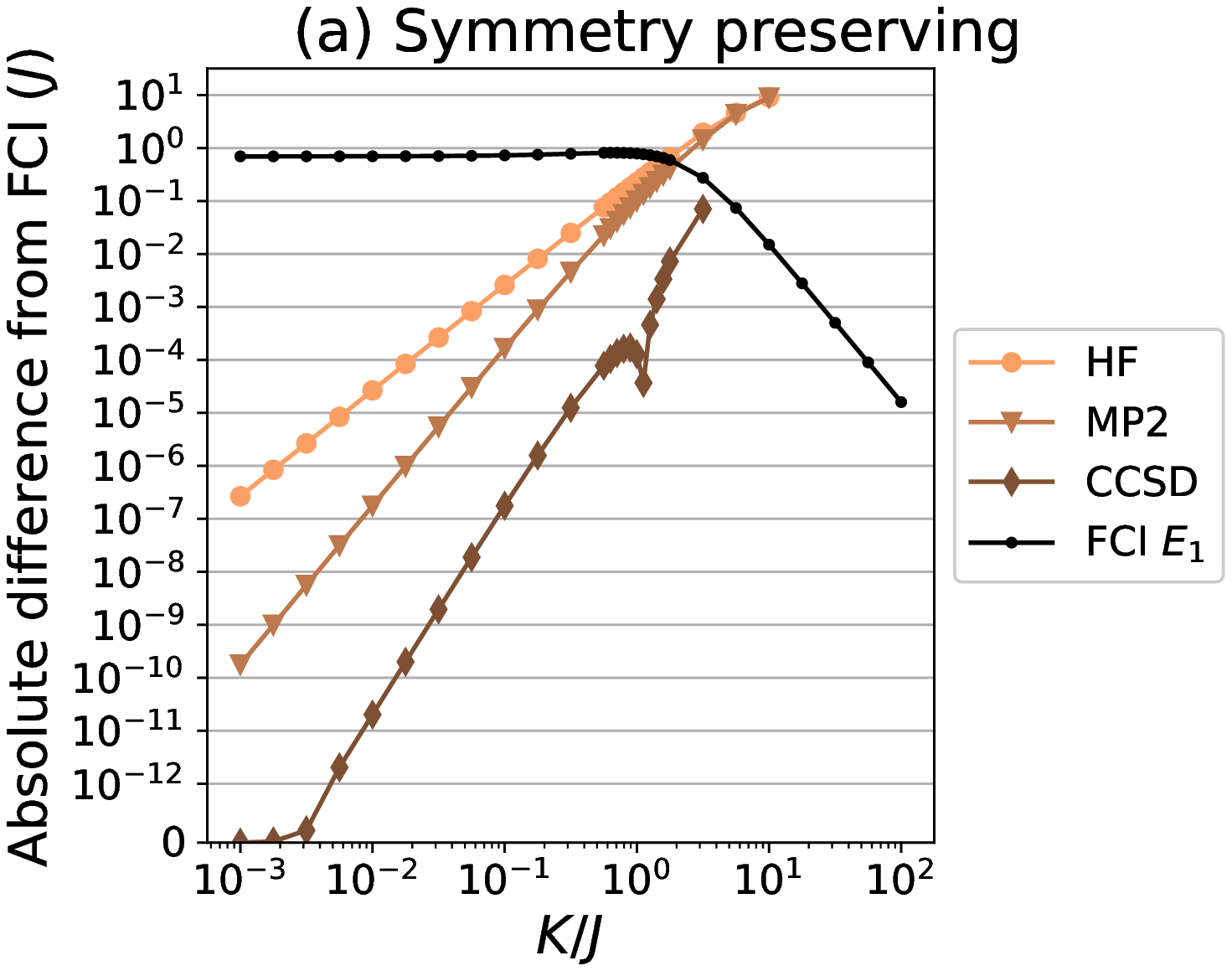}\label{fig:est_sym}
     \includegraphics[width=.45\linewidth]{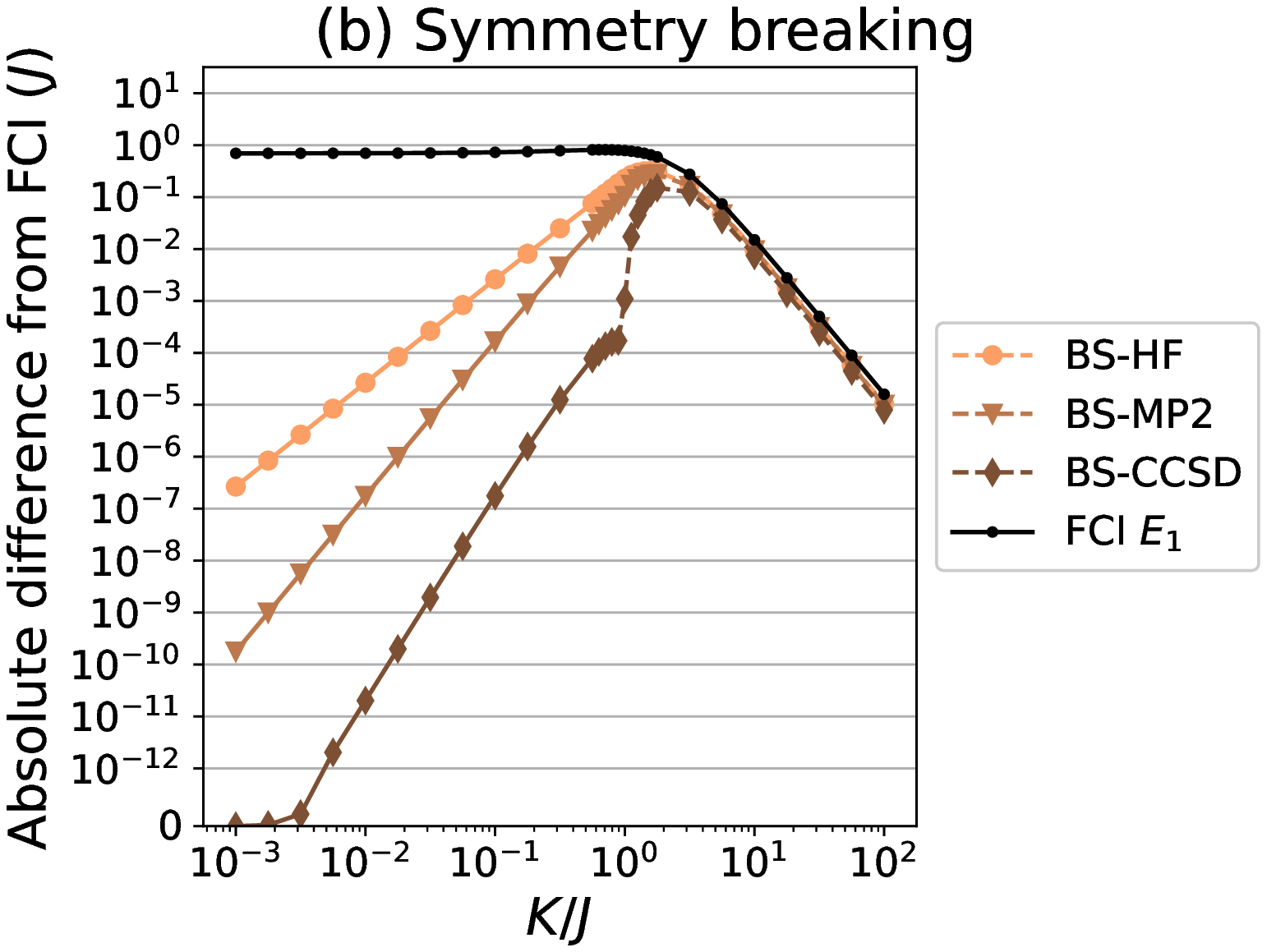}\label{fig:est_asym}
\caption{ \label{fig:est} Absolute energy errors from FCI for HF, MP2, and CCSD versus the correlation parameter $K/J$  presented alongside the energy gap between the ground and first excited FCI states (FCI $E_{1}$).
(a) Reference state is the symmetry preserving orbitals.
(b) Reference state is the broken-symmetry orbitals.
}
\end{figure*}

Fig.~\ref{fig:est}(a) presents the absolute errors for the symmetry-preserving HF, MP2 and CCSD on top of these references, and the energy gaps between the ground and first excited FCI states. 
For $K/J > 10$, the HF solutions begin to spontaneously break the parity symmetry of the system even when using the previous solutions at lower values of $K/J$ as initial guesses. 
We are also unable to converge the CCSD amplitude equations for $K/J > 3.16$. 
The kink in the CCSD data observed at $K/J = 1.12$ 
corresponds to the onset of CCSD having a lower energy than the FCI ground state. 
This non-variational behavior persists for all larger values of $K/J$. 

In the weakly correlated limit $K/J \ll 1$, MP2 reduces the energy error of HF by three orders of magnitude, demonstrating a success of simple perturbation theory. 
As more correlation is added to the system, the breakdown of MP2 becomes evident as its improvement over HF decreases significantly. 
For $K/J > 1$, the error of HF and MP2 continues to increase with increasing $K/J$. 
The HF and MP2 errors exceed the gap between the ground and first excited FCI states for $K/J > 1.78$ 
and $K/J > 3.16$, 
respectively. 
The error reduction of CCSD is more resilient to increasing correlation, though scanning from $K/J = 0.001$ to $K/J = 1$ the improvement in errors over HF reduces from over seven orders of magnitude to three orders of magnitude.


For $K/J \geq 1$, the symmetry-preserving HF solutions are found to be unstable to symmetry breaking. 
Fig.~\ref{fig:est}(b) presents the absolute errors for HF (BS-HF), MP2 (BS-MP2), and CCSD (BS-CCSD), where the HF solutions are allowed to break the parity symmetry of the lattice. 
The energy gap between the ground and first excited FCI states are also plotted. 
For $K/J < 1$, the HF solutions do not break symmetry and therefore the HF, MP2, and CCSD curves in this region are identical to those in Fig.~\ref{fig:est}(a). 
The cusp seen at $K/J = 1$ in the CCSD data arises from the change in character of the underlying HF reference due to symmetry breaking. 
Unlike in the symmetry-preserving case, the symmetry-broken CCSD remains above the FCI value for all surveyed values of $K/J$. 

In the symmetry-breaking regime, the errors in the HF, MP2, and CCSD solutions are much closer to one another compared to the $K/J < 1$ regime. 
CCSD improves upon the HF error by two orders of magnitude ($1.\times10^{-3}\text{ }J$ versus $2.\times10^{-1}\text{ }J$) for $K/J = 1$ while at $K/J = 100$, CCSD only improves on the HF error by ~20\% ($8.\times10^{-6}\text{ }J$ versus $1.\times10^{-5}\text{ }J$). 
For all $K/J$, the errors in the broken-symmetry HF, MP2, and CCSD results fall below the energy gap between the ground and first excited FCI states. 

The instability of HF to symmetry breaking for $K/J > 1$ is an example of the ``symmetry dilemma" discussed by L\"{o}wdin\cite{lowdin1955quantum}, wherein the most energetically favorable single determinant (classical state) breaks an intrinsic symmetry of the system, while the most energetically favorable symmetry-preserving state is higher in energy.  

\begin{figure*}
\caption{\label{fig:xxz_all}ADAPT-VQE results for the fermionized, anisotropic Heisenberg Hamiltonian. Absolute energy errors [(a), (d), and (g)], ADAPT-VQE gradient norms [(b), (e), and (h)], and infidelities from the exact wavefunction [(c), (f), and (i)] as the ADAPT-VQE ans\"{a}tze grow are presented for $K/J = 0.1$, $K/J = 1$, and $K/J = 10$. The ADAPT-VQE methods surveyed use symmetry-preserving HF orbitals and reference state (HF); broken symmetry HF orbitals and reference state (BS-HF); local site orbital basis, N\'{e}el reference state (Local, N\'{e}el); local site orbital basis, cat$_{+}$ reference state (Local, cat$_{+}$); and symmetry-adapted linear combination orbital basis, cat$_{+}$ reference state (SALC, cat$_{+}$). Absolute energy errors for the classical methods and the energy gap between the ground and first excited FCI states are presented alongside the ADAPT-VQE errors.}
\includegraphics[width=\linewidth]{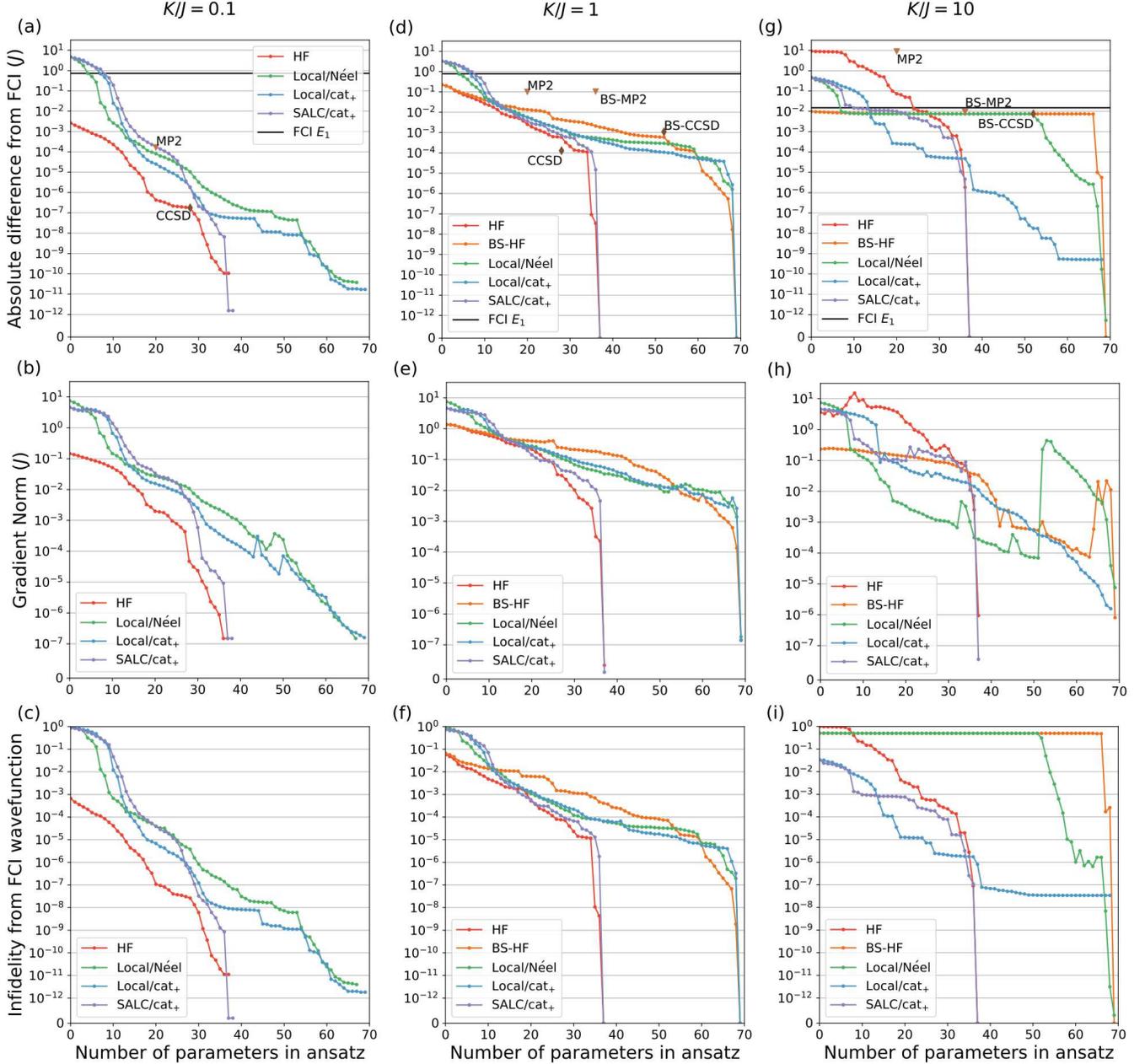}
\end{figure*}

Using the five calculation settings described above, we now investigate  the roles that parity symmetry and locality play in the compactness of the ADAPT-VQE ans{\"a}tze. 
In Fig.~\ref{fig:xxz_all},  we report the absolute error from the exact ground state energy, the norm of the ADAPT-VQE gradient, and the infidelity from the exact ground state wavefunction all versus the number of parameters in the ADAPT-VQE trial state. 
The infidelity, presented as a measure of closeness of the ADAPT-VQE wavefunction $|\psi^{\text{ADAPT}}\rangle$ to the exact ground state wavefunction $|\psi_{0}^{\text{FCI}}\rangle$, is given by
\begin{align}\label{eqn:inf}
    \text{Inf}\left(|\psi_{0}^{\text{FCI}}\rangle,|\psi^{\text{ADAPT}}\rangle\right) =& 1 - F\left(|\psi_{0}^{\text{FCI}}\rangle,|\psi^{\text{ADAPT}}\rangle\right) \\
    =& 1 - \left|\langle\psi_{0}^{\text{FCI}}|\psi^{\text{ADAPT}}\rangle\right|^{2},
\end{align}
where $F$ is the fidelity of the two states.

\subsubsection{Symmetry breaking slows energy convergence}
Looking across panels \ref{fig:xxz_all}(a), \ref{fig:xxz_all}(d), and \ref{fig:xxz_all}(g),
we see that in each case, both symmetry preserving calculations (HF and SALC/cat$_+$)
converged to the exact solution with only 37 parameters. 
This reduced number of parameters arises from the fact that when symmetry is preserved throughout the state preparation, 
we only need to parameterize states within the corresponding $g$-symmetry subspace, which for this 8-site lattice has a dimension of 38.
For all the other cases, which involve some symmetry breaking (either in the orbitals, reference state, or both),
the full 70-dimensional Hilbert space must be spanned to achieve exact convergence. 
The 52 CCSD amplitudes can be divided into 28 $g$ excitations and 24 $u$ excitations.
In Fig. \ref{fig:xxz_all}, we only report CCSD as having 28 parameters, since the remaining 24 $u$-symmetry excitations cannot contribute when applied to a symmetry-preserving reference state. 
As such, CCSD does not have enough parameters to span the 38-dimensional $g$ subspace. 
This is also obvious from the fact that CCSD has neither connected triple nor quadruple excitations, 
interactions that ADAPT-VQE is able to include by sequential application of one- and two-particle rotations. 


For the weaker-correlation case ($K/J = 0.1$),  
the symmetry-preserving calculations always outperform the symmetry-violating cases (the red curve is always below the rest).
This is not too surprising, 
given that the HF reference state is the most stable product state available. 
With the onset of symmetry-breaking in HF ($K/J = 1$), the broken-symmetry HF is slightly more favorable. However, ADAPT(HF) begins to outperform ADAPT(BS-HF) after a single iteration. 
On the other hand, when the correlation increases to $K/J=10$, 
the symmetry-preserving HF reference is no longer the lowest-energy product state, but instead is now the highest-energy reference state considered in our data. 
As a consequence, in this strong-correlation regime, the use of a broken-symmetry reference leads to lower energy at early stages of the algorithm. 
However, this energetic advantage of the broken-symmetry reference quickly becomes a disadvantage due to a very slow convergence at later stages [seen as the flat-lining of the green and orange curves in Fig.~\ref{fig:xxz_all}(g)]. 

 \subsubsection{Symmetry breaking worsens gradient troughs}
As reported recently,\cite{grimsleyADAPTVQEInsensitiveRough2022} strongly correlated systems are susceptible to exhibiting gradient troughs, 
whereby the gradients of the pool operators initially diminish before eventually increasing prior to convergence. This non-monotonic convergence is problematic because it appears to the user as false convergence. 
For the fermionized, anisotropic Heisenberg model studied here, we again observe the onset of gradient troughs when the correlation is increased,
as is clearly evident in Fig.~\ref{fig:xxz_all}(h).
However, when we allow the symmetry to break, we find that problems with gradient troughs worsen.

For both symmetry-breaking references,  ADAPT(BS-HF) (orange) and ADAPT(local/N\'{e}el) (green), the norm of the operator pool gradient is seen to decrease with the addition of operators to the ansatz and then suddenly jump by several orders of magnitude. 
Before escaping from the gradient trough, the energy errors for ADAPT(BS-HF) and ADAPT(local/N\'{e}el) lie close to that of broken-symmetry CCSD, which is approximately one half of the gap between the exact ground and first excited FCI states [panel ~\ref{fig:xxz_all}(g)]. 
Additionally, the infidelities of the ADAPT(BS-HF) and ADAPT(local/N\'{e}el) states are approximately 0.5 before escaping the gradient trough [panel ~\ref{fig:xxz_all}(i)]. 
In these cases, broken-symmetry CCSD, ADAPT(BS-HF), and ADAPT(local/N\'{e}el) appear to be approximating a broken-symmetry state which is an equal superposition of the ground and first-excited states:
\begin{equation}\label{eqn:bs_wfn}
    |\psi^{\text{BS}}\rangle = \frac{1}{\sqrt{2}}\left(|\psi_{0}^{\text{FCI}}\rangle + |\psi_{1}^{\text{FCI}}\rangle\right).
\end{equation}
The energy error associated with this state is 
\begin{eqnarray}\label{eqn:bs_energy}
    \Delta E^{\text{BS}} &=& \langle\psi^{\text{BS}}|\hat{H}|\psi^{\text{BS}}\rangle - E_{0} \nonumber \\
    &=& \frac{1}{2}\left(E_{1} + E_{0}\right) - E_{0} \\
    &=& \frac{1}{2}\left(E_{1}-E_{0}\right) \nonumber
\end{eqnarray}
and the infidelity of this state is
\begin{eqnarray}
    \text{Infidelity}\left(|\psi_{0}^{\text{FCI}}\rangle,|\psi^{\text{BS}}\rangle\right) &=& 1 - \left|\langle\psi_{0}^{\text{FCI}}|\psi^{\text{BS}}\rangle\right|^{2} \nonumber \\
    &=& \frac{1}{2}. 
\end{eqnarray}

To explain this behavior, consider the N\'{e}el reference state $|10101010\rangle$ and its complement $|01010101\rangle$. 
The exact solution has equal weights for these states.%
\footnote{Although asymptotically large values of $K/J$ admit arbitary mixtures of these two states,
any finite value will contributions from other configurations which will fix the relative weight between these configurations to be equal.}
As $K/J$ becomes large, the Hamiltonian more strongly penalizes states with occupation on consecutive sites. 
Because the operator pool includes only single and double excitations, a single operator cannot enact the quadruple excitation required to go between the reference state and its complement. 
The weight of the complement state in the ADAPT-VQE wavefunction therefore is generated via products of multiple lower-rank excitation operators, putting it out of reach for a single pool operator. 
ADAPT-VQE(local/N\'{e}el) first touches the complement N\'{e}el state after four operator additions. 
Despite having access to this determinant, ADAPT-VQE does not have the variational flexibility to significantly weigh this state, as doing so would consequently weigh higher-energy intermediate determinants, raising the energy. 
As ADAPT-VQE continues to add operators, additional excitation pathways begin to form, though the VQE subroutine keeps the weight on the complementary state small. 
With the addition of the 52nd operator, ADAPT-VQE achieves the variational flexibility to substantially increase the weight of the complementary N\'{e}el state. 
This significant change in the character of the ADAPT-VQE state is reflected in subplots \ref{fig:xxz_all}(g), \ref{fig:xxz_all}(h), and \ref{fig:xxz_all}(i): the energy begins to significantly decrease again, the gradient norm jumps, and the infidelity drops. 

The suppression of these pathways leads to a deeper gradient trough with increasing $K/J$. 
To further explain this suppression of the operator gradient, we consider Eq.~\ref{eqn:adapt_grad} with the ADAPT-VQE state expressed in terms of the eigenstates $|\psi^{\text{FCI}}_{i}\rangle$ of $\hat{H}$:
\begin{align}
    |\psi^{(n)}\rangle =& \sum_{i}c^{(n)}_{i}|\psi^{\text{FCI}}_{i}\rangle \\
    \frac{\partial E}{\partial \theta_{k}} =& \sum_{i}\sum_{j}c^{(n)*}_{i}c^{(n)}_{j}\left\langle\psi^{\text{FCI}}_{i}\left|\left[\hat{H},\hat{A}_{k}\right]\right|\psi^{\text{FCI}}_{j}\right\rangle \\
    =& \sum_{ij}c^{(n)*}_{i}c^{(n)}_{j}(E_{i}-E_{j})\left\langle \psi^{\rm{FCI}}_{i}\left|\hat{A}_{k}\right|\psi^{\rm{FCI}}_{j}\right\rangle. \label{eqn:adapt_grad2}
\end{align}
The energy difference term here is seen to suppress the gradients when a contaminant state and the target state become close in energy. 
As $K/J$ becomes large, the gap between the exact ground and first excited states shrinks, suppressing the gradients in this regime when the first excited state is a major contaminant in the ADAPT-VQE trial state, as in the cases with broken-symmetry reference states. 

For even stronger correlation ($K/J = 100$), ADAPT(BS-HF) and ADAPT(local/N\'{e}el) become fatally trapped in a gradient trough (see Supplementary Information).  
This can be seen as suppression of the operator gradient below the tolerance of the numerical noise of the VQE optimizer. 
As such, these methods with broken-symmetry reference states retain high infidelities and energy errors. 
We speculate that, with a numerically exact optimizer, ADAPT-VQE should be able to escape even these gradient troughs,
although this would not be possible for a quantum computer with finite noise. 

The emergence of deep gradient troughs is not seen for ADAPT(local/cat$_{+}$), even for large $K/J$. 
ADAPT(HF) exhibits a shallow gradient trough at the start of the ADAPT-VQE procedure. 
In this case, the $g$-symmetry reference state is a superposition of the ground state and excited states with $g$-symmetry. 
The high infidelity (0.98) of the initial state indicates severe contamination. 
The symmetry of the operator pool made from symmetry-preserving HF orbitals, however, prevents contamination from the low-lying, $u$-symmetry first excited state. 
This restriction is seen to limit the depth of the trough. 
These results highlight the importance of symmetry in avoiding these deep gradient troughs. 


\subsection{\ce{H4}}
\begin{figure*}
{\includegraphics[width=.49\linewidth]{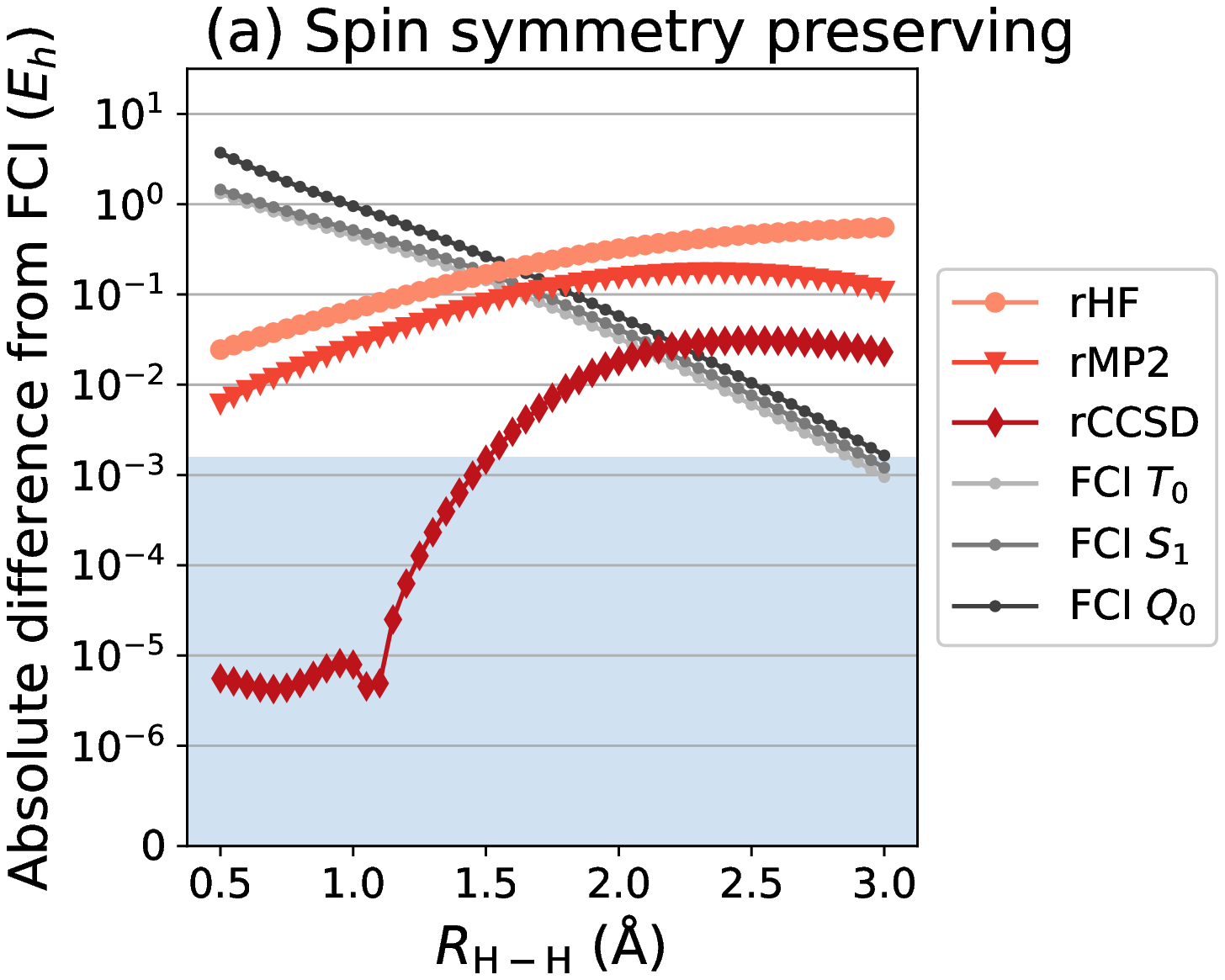}\label{fig:h4_est_rhf}} 
{\includegraphics[width=.49\linewidth]{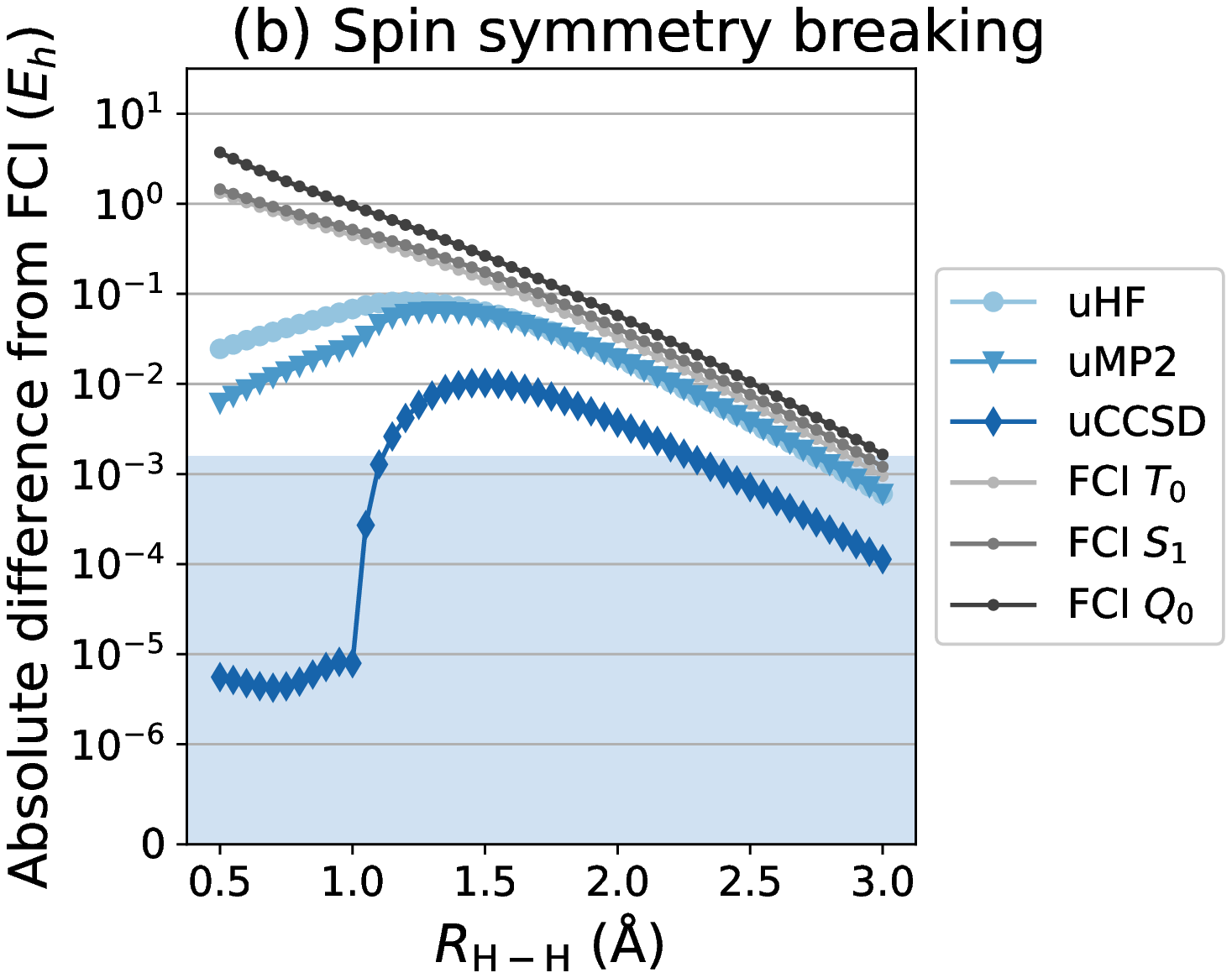}\label{fig:h4_est_uhf}} 
\caption{\label{fig:h4_est} Absolute energy errors from FCI for HF, MP2, and CCSD versus the \ce{H}-\ce{H} separation presented alongside the energy gaps between the ground and next lowest singlet ($S_{1}$), triplet ($T_{0}$), and quintet ($Q_{0}$) excited states. HF spontaneously breaks $\hat{S}^{2}$-symmetry for $R_{\text{\ce{H}-\ce{H}}} > 1.00\,\text{\AA}$. The shaded area denotes ``chemical accuracy" (errors less than one kcal mol$^{-1}$).
(a) Reference state is restricted HF (rHF).
(b) Reference state is unrestricted HF (uHF).}
\end{figure*}

In Fig.~\ref{fig:h4_est} we present the absolute errors for a series of ``classical'' quantum chemistry methods.
Fig.~\ref{fig:h4_est}(a) presents restricted HF (rHF), MP2 and CCSD on top of this reference (rMP2 and rCCSD, respectively), and the energy gaps between the FCI singlet ground state and the next lowest singlet ($S_{1}$), triplet ($T_{0}$), and quintet ($Q_{0}$) FCI excited states for the dissociation of \ce{H4}. 
These excited states become degenerate with the ground state in the dissociation limit.  
CCSD yields a lower energy than the FCI ground state for values of $R_{\text{\ce{H}-\ce{H}}} > 1.05\,\text{\AA}$, yielding a kink in the absolute energy errors. 

For $R_{\text{\ce{H}-\ce{H}}} = 0.90\,\text{\AA}$, near the equilibrium geometry, rMP2 improves upon the rHF reference error by more than a factor of two ($2.\times10^{-2}\,E_{h}$ versus $6.\times10^{-2}\,E_{h}$), while rCCSD improves upon the reference by nearly four orders of magnitude. 
This improvement of rMP2 over rHF increases near the dissociation limit as the rMP2 energy begins to ``turn over", with the rMP2 energy peaking at $R_{\text{\ce{H}-\ce{H}}} = 2.45\,\text{\AA}$. 
Beginning at the onset of the non-variational behavior, the improvement of rCCSD over rHF absolute errors decreases with increasing \ce{H}--\ce{H} separation to just over one order of magnitude ($-0.03\,E_{h}$ versus $0.5\,E_{h}$) at $R_{\text{\ce{H}-\ce{H}}} = 2.50\,\text{\AA}$, at which point the rCCSD energy begins to ``turn up". 
The rCCSD energies are seen to be ``chemically accurate" (errors less than one kcal mol$^{-1}$) for $R_{\text{\ce{H}-\ce{H}}} \leq 1.50\,\text{\AA}$. 
For large \ce{H}--\ce{H} separations, the rHF, rMP2, and rCCSD absolute errors exceed the energy gaps between the FCI ground state and the $T_{0}$, $S_{1}$, and $Q_{0}$ states. 

Fig.~\ref{fig:h4_est}(b) presents the absolute errors for unrestricted HF (uHF), MP2 and CCSD on top of this reference (uMP2 and uCCSD, respectively), and the energy gaps between the FCI singlet ground state and the next lowest singlet ($S_{1}$), triplet ($T_{0}$), and quintet ($Q_{0}$) FCI excited states for the dissociation of \ce{H4}. 
The uHF references spontaneously break $\hat{S}^{2}$ symmetry for $R_{\text{\ce{H}-\ce{H}}} > 1.00\,\text{\AA}$. 
The kinks observed in the uMP2 and uCCSD errors at these points correspond to the change in the underlying HF reference. 
The uCCSD energy errors remain positive for all $R_{\text{\ce{H}-\ce{H}}}$ surveyed. 

Unlike in the restricted case, the improvement of uMP2 over uHF becomes negligible as the \ce{H}--\ce{H} separation increases. 
The improvement of uCCSD over uHF decreases from over two orders of magnitude at the onset of symmetry breaking ($R_{\text{\ce{H}-\ce{H}}} = 1.05\,\text{\AA}$; $3.\times10^{-4}\,E_{h}$ versus $7.\times10^{-2}\,E_{h}$) to a factor of five at $R_{\text{\ce{H}-\ce{H}}} =3.00\,\text{\AA}$ ($6.\times10^{-4}\,E_{h}$ versus $1.\times10^{-4}\,E_{h}$). 
The uCCSD energies are chemically accurate for $R_{\text{\ce{H}-\ce{H}}} \leq 1.10\,\text{\AA}$ and $R_{\text{\ce{H}-\ce{H}}} \geq 2.30\,\text{\AA}$. 
Both uHF and uMP2 energies are chemically accurate for $R_{\text{\ce{H}-\ce{H}}} \geq 2.75\,\text{\AA}$. 
The errors for uHF, uMP2, and uCCSD fall below the energy gaps between the FCI ground and excited states for all \ce{H}--\ce{H} separations investigated. 

\begin{figure*}
\includegraphics[width=\linewidth]{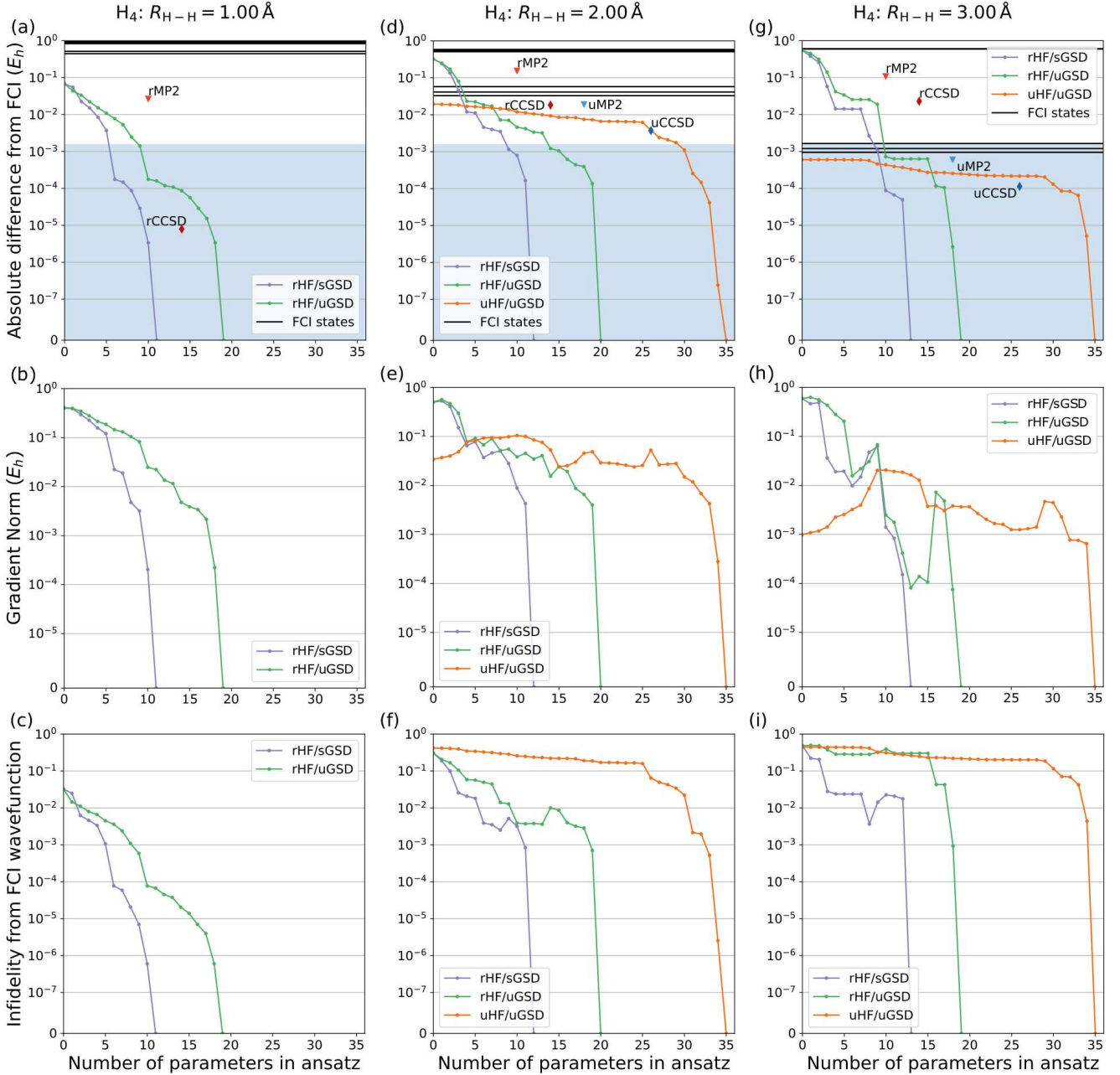}
\caption{\label{fig:all_h4} ADAPT-VQE results for the symmetric dissociation of linear \ce{H4}. Absolute energy errors ((a), (d), and (g)), ADAPT-VQE gradient norms [(b), (e), and (h)], and infidelities relative to the exact wavefunction ((c), (f), and (i)) as the ADAPT-VQE ans{\"a}tze grow are presented for $R_{\text{\ce{H}-\ce{H}}} = 1.00\text{\AA}$, $R_{\text{\ce{H}-\ce{H}}} = 2.00\text{\AA}$, and $R_{\text{\ce{H}-\ce{H}}} = 3.00\text{\AA}$. The ADAPT-VQE methods with restricted HF orbitals, reference states, and singlet GSD operator pools (rHF/sGSD);  restricted HF orbitals, reference states, and unrestricted GSD pools (rHF/uGSD); and unrestricted HF orbitals, reference states, and unrestricted GSD operator pools (uHF/uGSD) are presented. Absolute energy errors for the classical methods and the energy gaps between the ground state and the 9 lowest excited states are presented alongside the ADAPT-VQE errors.}
\end{figure*}

We now investigate the role that spin symmetry plays in the compactness of the ADAPT-VQE ans\"{a}tze for the symmetric dissociation of linear \ce{H4}. 
Fig.~\ref{fig:all_h4} presents the absolute energy errors ((a), (d), and (g)), ADAPT-VQE gradient norms ((b), (e), and (h)), and infidelities from the exact wavefunction ((c), (f), and (i)) as the ans\"{a}tze grow for ADAPT-VQE applied to the symmetric dissociation of \ce{H4} at $R_{\text{\ce{H}-\ce{H}}} = 1.00\,\text{\AA}$, $R_{\text{\ce{H}-\ce{H}}} = 2.00\text{\AA}$, and $R_{\text{\ce{H}-\ce{H}}} = 3.00\text{\AA}$. 
The ADAPT-VQE methods surveyed are ADAPT-VQE using rHF orbitals, rHF reference states, and singlet GSD operator pools [ADAPT(rHF/sGSD)]; ADAPT-VQE using rHF orbitals, rHF reference states, and unrestricted GSD operator pools [ADAPT(rHF/uGSD)], and ADAPT-VQE using uHF orbitals, uHF reference states, and unrestricted GSD operator pools [ADAPT(uHF/uGSD)].   

\subsubsection{Spin symmetry breaking slows energy convergence}
Comparing panels \ref{fig:all_h4}(a), \ref{fig:all_h4}(d), and \ref{fig:all_h4}(g), we see that ADAPT(rHF/sGSD) converges to the exact solution with 11, 12, and 13 parameters for $R_{\text{\ce{H}-\ce{H}}} = 1.00\text{\AA}$, $R_{\text{\ce{H}-\ce{H}}} = 2.00\text{\AA}$, and $R_{\text{\ce{H}-\ce{H}}} = 3.00\text{\AA}$, respectively.  
The use of the sGSD operator pool ensures the ADAPT-VQE state remains an eigenstate of $\hat{S}^{2}$. 
While there are 20 determinants in the basis of rHF orbitals that contribute to the exact ground state, by enforcing spin-symmetry, ADAPT(rHF/sGSD) is able to converge to the exact solution with as few as 11 parameters. 
For $R_{\text{\ce{H}-\ce{H}}} = 2.00\text{\AA}$ and $R_{\text{\ce{H}-\ce{H}}} = 3.00\text{\AA}$, ADAPT(rHF/sGSD) converges to local minima when 11 operators have been added, and the addition of one and two additional operators, respectively, increases the variational flexibility of the ADAPT-VQE state and allows it to recover the global minimum. 
For ADAPT(rHF/uGSD), the uGSD operator pool contains operators that break $\hat{S}^{2}$ symmetry.
Therefore despite beginning with the rHF reference state, as the ADAPT(rHF/uGSD) ansatz grows the $\langle\hat{S}^{2}\rangle$ expectation value is seen to deviate from 0. 
Without the efficient parameterization offered by the sGSD operator pool, ADAPT(rHF/uGSD) requires at least 19 parameters to converge to the exact solution of the 20-dimensional subspace. 
For ADAPT(uHF/uGSD), the use of uHF orbitals breaks the symmetries between the $\alpha$ and $\beta$ orbitals, and thus there are 36 unrestricted determinants (corresponding to all $\hat{S}_{z}$-preserving determinants) that contribute to the exact ground state. 
In this case the full 36-dimensional Hilbert subspace must be spanned to achieve convergence to the exact ground state, requiring at least 35 parameters. 

Comparing ADAPT(rHF/sGSD) and ADAPT(rHF/uGSD), the use of the $\hat{S}^{2}$-preserving operator pool not only accelerates convergence to the exact ground state but is also seen to require fewer parameters to achieve chemical accuracy for all \ce{H}-\ce{H} separations surveyed. 
After the onset of symmetry-breaking in the HF reference state (beginning near $R_{\text{\ce{H}-\ce{H}}} = 1.05\,\text{\AA}$), ADAPT(uHF/uGSD) initially outperforms ADAPT(rHF/sGSD) and ADAPT(rHF/uGSD) by virtue of a more energetically favorable reference state.
Despite this, the more efficient parameterization offered by preserving symmetries allows both ADAPT(rHF/sGSD) and ADAPT(rHF/uGSD) to outperform ADAPT(uHF/uGSD) after the addition of only a few operators. 
For $R_{\text{\ce{H}-\ce{H}}} = 2.00\,\text{\AA}$, these crossovers occur before ADAPT(uHF/uGSD) has achieved chemical accuracy, whereas for $R_{\text{\ce{H}-\ce{H}}} = 3.00\,\text{\AA}$, ADAPT(uHF/uGSD) achieves chemical accuracy before ADAPT(rHF/sGSD) and ADAPT(rHF/uGSD), as the uHF reference state is already chemically accurate. 
Despite this, ADAPT(uHF/uGSD) shows very limited improvement in error as more operators are added, and it has not significantly improved upon the reference energy when the crossovers with ADAPT(rHF/sGSD) and ADAPT(rHF/uGSD) are reached. 

For $R_{\text{\ce{H}-\ce{H}}} = 1.00\,\text{\AA}$, the rHF reference state provides a reasonable zeroth order description of the system, and as such rCCSD is able to provide a competitive performance to ADAPT(rHF/uGSD) with the same number of parameters despite the lack of connected triple and quadruple excitations. 
As the \ce{H}-\ce{H} separation increases, the rHF reference state becomes a poorer description of the true ground state, as seen by the initial infidelities [panels \ref{fig:all_h4}(c), \ref{fig:all_h4}(f), and \ref{fig:all_h4}(i)]. 
Here these excitations become more important in accurately describing the ansatz, and as such the performance of rCCSD is seen to suffer relative to ADAPT(rHF/uGSD) for the same number of parameters. 

\subsubsection{Spin symmetry breaking worsens gradient troughs}
For large \ce{H}-\ce{H} separations, all three of the ADAPT-VQE methods surveyed exhibit gradient troughs, as is evident in Fig.~\ref{fig:all_h4}(h). 
These gradient troughs are accompanied by a flattening of the energy error curve [Fig.~\ref{fig:all_h4}(g)] and large infidelities [Fig.~\ref{fig:all_h4}(i)]. 

The rHF and uHF reference states at this geometry have high infidelities with respect to the exact ground state, (0.482 and 0.442, respectively), indicating that these trial ground states contain significant contributions from excited FCI states. 
Additionally, the lowest singlet ($S_{1}$), lowest triplet ($T_{0}$), and lowest quintet ($Q_{0}$) excited states lie close in energy to the ground state. 
Recalling Eq.~\eqref{eqn:adapt_grad2}, we see that the ADAPT-VQE gradients are suppressed when the overlap between the ADAPT-VQE state and the target state are small (small $c^{(n)}_{0}$) and when the contaminant states are close in energy to the target state [small ($E_{j} - E_{0}$)]. 


For ADAPT(rHF/sGSD), the rHF reference state is a singlet and as such all states contributing to it are singlet in nature. 
By enforcing the ADAPT-VQE trial state to be a singlet, the effect of the sGSD operator pool is to limit the possible contaminant states. 
This results in a gradient trough that is relatively shallow, and ADAPT(rHF/sGSD) acquires the variational flexibility to escape the trough by adding only a few operators. 
While ADAPT(rHF/uGSD) utilizes the same singlet rHF reference state as ADAPT(rHF/sGSD) and as such begins with only singlet contaminant states, the use of the uGSD operator pool introduces contaminant states of higher spin multiplicities as the ADAPT-VQE procedure proceeds, beginning with the second operator addition. 
This is evidenced by the initial growth of the $\langle\hat{S}^{2}\rangle$ expectation value, and can be understood as a variational conversion of higher-energy, singlet contaminant states to lower-energy, higher-spin-multiplet states. 
The uHF reference state has an $\langle\hat{S}^{2}\rangle$ expectation value of 1.996, indicating significant contamination from the $T_{0}$ and $Q_{0}$ excited states.
ADAPT(rHF/uGSD) and ADAPT(uHF/uGSD) are both seen to exhibit two gradient troughs. 
Escaping each of these gradient troughs is accompanied by a drop in the $\langle\hat{S}^{2}\rangle$ expectation value. 
In the case of ADAPT(uHF/uGSD), this can be understood as ADAPT-VQE acquiring the variational flexibility to project out contamination from $Q_{0}$, corresponding to a drop in $\langle\hat{S}^{2}\rangle$ from $\sim$2 to $\sim$0.6, and subsequently acquiring the variational flexibility to project out contamination from $T_{0}$, corresponding to a drop in $\langle\hat{S}^{2}\rangle$ from $\sim$0.6 to 0 at convergence to the exact ground state. 

\section{Conclusions}
In this work, we have investigated two strongly correlated systems that exhibit two different kinds of spontaneous symmetry breaking at the mean-field level as correlation increases. 
In each case, we explore the role that breaking/preserving these symmetries in the reference states, operator pools, and representations of the Hamiltonian has on the performance of ADAPT-VQE. 

While reducing symmetry through the use of UHF orbitals often improves the energy accuracy of classical electronic structure theory methods, the use of broken-symmetry HF solutions is a detriment to ADAPT-VQE. 
For fermionic operator pools without symmetry-adaptation of the operators, the symmetry (or lack thereof) of the pools is determined by the symmetries of underlying orbitals. 
With the onset of symmetry breaking in the MO basis, the number of determinants with nonzero weights in the expansion of the exact ground state increases significantly.
In order to create the exact ground state, each of the determinants contributing to the exact ground state requires the addition of an operator to the ansatz. 
Thus, the use of symmetry-broken HF as a reference for ADAPT-VQE, though improving the energy of the reference, leads to much larger exact ans\"{a}tze compared to symmetry-preserving HF/rHF.

In the local representation of the Hamiltonian, the underlying site orbital basis is inherently symmetry-broken, and as such the representation of the exact ground state in the determinant basis made of the site orbitals is dense. 
For these systems, the use of operator pools that are not symmetry-adapted again requires a larger number of operators to converge ADAPT-VQE. 
This is the case whether the reference state is symmetrized (cat$_{+}$) or broken symmetry (N\'{e}el).

Symmetries can be introduced to the operator pool by changing the underlying orbital basis (SALC) or via symmetry-adaptation of the pool operators (using the sGSD pool for \ce{H4}). 
In both cases, the preservation of these symmetries leads to shorter ans\"atze with ADAPT-VQE. 
In the former, transformation of the site basis yields a sparser representation of the exact ground state in the new orbital basis, and thus an operator pool without symmetry-adaptation using this orbital basis leads to ADAPT-VQE convergence with a smaller number of operators compared to the original site orbital basis.
In the latter, the singlet GSD pool more efficiently spans the subspace of determinants that overlap with the exact ground state by parameterizing a symmetry-adapted combination of fermionic operators with a single parameter. 

With respect to the issue of gradient troughs in ADAPT-VQE, we make the following observations:
\begin{enumerate}
    \item Gradient troughs appear when excited states become close in energy to the ground state, such as the large $K/J$ limit of the fermionized, aniostropic Heisenberg model or in the limit of large \ce{H}--\ce{H} separation in linear \ce{H4}. 
    \item Reference states that have a high fidelity with the exact ground state do not exhibit deep gradient troughs (cat$_{+}$), while the reference states with low fidelities are seen to exhibit them when low-lying excited states are present. 
    \item For systems where the reference state is symmetry-preserving, the use of symmetry-adapted operator pools leads to shallow troughs [ADAPT(HF), ADAPT(rHF/sGSD)], while symmetry-agnostic operator pools can lead to deep gradient troughs when the overlap with the exact ground state is low [ADAPT(rHF/uGSD)]. Using symmetry-adapted operator pools limits the possible contaminant states in the ADAPT-VQE state to those that obey the symmetry in question, while symmetry-agnostic pools may introduce new contaminants into trial states that were not initially present.
    \item For symmetry-broken reference states, the presence of deep gradient troughs is endemic (ADAPT(BS-HF), ADAPT(local/N\'{e}el), ADAPT(uHF/uGSD)). 
\end{enumerate}

While preparing this manuscript for publication,  a relevant preprint  by Tsuchimochi et al\cite{tsuchimochi2022adaptive} appeared that also looks at spin-symmetry breaking in ADAPT-VQE. In their work they highlight the unfavorable behavior of the ``spin-dependent fermionic operator pool" (unrestricted operator pools in this work) and spin-complemented operator pools to break $\hat{S}^{2}$ symmetry. The authors similarly find that spin-symmetry breaking leads to an increase in the quantum computational resources (both parameter counts and CNOT gates required) compared to their spin-projected ADAPT-VQE, which applies a spin projection operator to restore the $\hat{S}^{2}$ symmetry. They further apply this spin-projected ADAPT-VQE to the computation of molecular properties and geometry optimization. 

\begin{acknowledgments}
This work was supported by the National Science Foundation (Award No. 1839136). The authors thank Advanced Resource Computing at Virginia Tech for use of computational resources. L.W.B thanks Dr. Ayush Asthana  and Dr. John Van Dyke for useful discussions relevant to the work. 
\end{acknowledgments}

\bibliography{bibliography, nick}

\end{document}